\documentclass[preprint,12pt]{elsarticle}

\usepackage{graphicx}
\usepackage{amssymb}
\usepackage{color}

\newcommand{\blue}{\textcolor{black}}

\journal{Appl. Surf. Sci.}

\begin{document}

\begin{frontmatter}

\bibliographystyle{elsarticle-num}


\title{Paracetamol Micro-Structure Analysis by Optical Mapping}


\author{Reo Honda$^1$, Meguya Ryu$^1$, Armandas Bal\v{c}ytis$^2$, Jitraporn Vongsvivut$^3$, Mark J. Tobin$^3$, Saulius Juodkazis$^{2,4}$, Junko Morikawa$^1$\cortext[mycorrespondingauthor]}
\address{$^1$~Tokyo Institute of Technology, Meguro-ku, Tokyo 152-8550, Japan \\\protect $^2$ Nanotechnology facility, Swinburne University of
Technology, John st., Hawthorn, 3122 Vic, Australia \\\protect $^3$ Infrared Microspectroscopy Beamline, Australian Synchrotron, Clayton, 3168 Vic, Australia \\\protect $^4$ Melbourne Center for Nanofabrication, Australian National Fabrication Facility, Clayton, 3168 Vic, Australia}


\cortext[mycorrespondingauthor]{Corresponding author: Junko Morikawa morikawa.j.aa@m.titech.ac.jp }

\begin{abstract}
Domain structure of paracetamol - popular antipyretic analgesic - was investigated by infrared (IR) spectroscopy using synchrotron radiation. Absorbance and retardance maps reveal molecular orientation inside the micro-domains of the paracetamol form II which has a better water solubility and compressibility compared to the commercially used forms I. The developed method of analysis representing orientation of optical slow(fast)-axis is compared with azimuthal orientation of the absorbance at several specific IR bands using vector maps. High brightness of synchrotron radiation and hyper-spectral mapping of structural domains in paracetamol clearly reveals the domain boundaries and can potentially be used to observe \emph{in situ} intra-phase transformations of paracetamol forms-I, II, III and melting, which are important for making medical tablets and powders by an industrial process.  
\end{abstract}
\begin{keyword}
\texttt{orientational anisotropy\sep absorbance\sep retardance\sep paracetamol\sep  water solubility}
\end{keyword}
\end{frontmatter}
\newpage
\section{Introduction}

Making pressed tablets, encapsulation of pharmaceuticals for a controlled release, water solubility of drinkable fizzing medications, are \blue{all} based on the state-of-the-art research in packaging with new developments guided by active \blue{pharmaceutical}  markets~\cite{Kawashima,Clusel}. Water solubility, thermal stability of medically active phases, their mixing, compressibility, and mechanical structure are the physical attributes \blue{driven} by \blue{a number of} multi-disciplinary approaches~\cite{Liang,Espinosa}. 

Paracetamol is a pharmaceutical used for its antipyretic and analgesic properties. It has three polymorphs: forms I~\cite{Haisa}, II~\cite{parac}, III~\cite{Perrin} at atmospheric pressure. The form I is the most stable and it is used as pharmaceutical material. The forms II and III are \blue{metastable phases, which can} undergo solid phase transitions. \blue{Water solubility is of a key importance for the medical applications. The metastable phases and amorphous  paracetamol are water soluble and the water solubility is higher than that of the form I, which is used in the industrial production.} \blue{Insights into how} the molecular orientation and domain structure alters water solubility  of the very same chemical compound and how it can be altered by phase transitions, \blue{are crucially important for enhancing the efficacy of the drugs}. Distribution and orientation of OH groups on the surface and along the grain boundaries can help to elucidate intricacies of solubility.  The orientation of the chemical bonds is the most fundamental information of the ordered structure of the bio-active molecular materials. The form I of paracetamol is a stable polymorph used commercially to make tablets. However, the form II crystal investigated here has a higher water solubility and compressibility advantageous for pharmaceutical applications~\cite{Martino}.  

\begin{figure}[t!]
\begin{center}
\includegraphics[width=13cm]{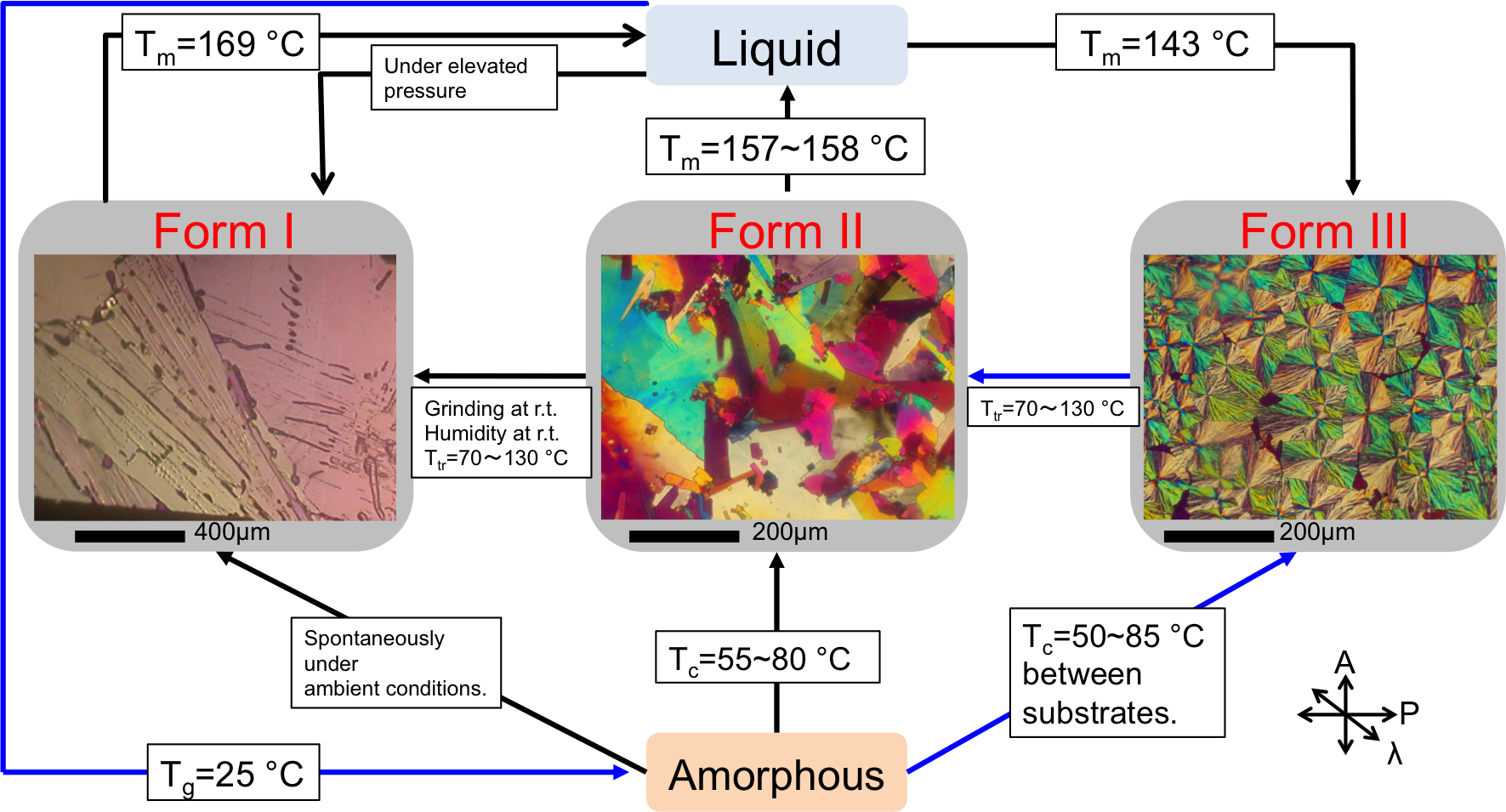}
\caption{Polarized optical microscope (POM) images of three polymorphs of paracetamol (forms I, II, III)~\cite{Haisa,parac,Perrin} and possible phase transitions; \blue{$T_{g,m,tr}$ are temperatures of glass transition, melting and solid phase transition, respectively~\cite{Martino,Gaisford,Qi,Martino2,John,Kach,Zimm}. Blue arrows show the preparation procedure of the sample used for IR measurement in this study.} The polariser (P) and analyser (A) orientations are shown together with that of a waveplate $\lambda = 530$~nm used to color-shift the polariscopy image. Thickness of paracetamol film was $d\approx 10~\mu$m for the forms II and III, while $\sim 25~\mu$m for the form I. }
\label{f-forms}
\end{center}
\end{figure}

\begin{figure}[t!]
\begin{center}
\includegraphics[width=8.5cm]{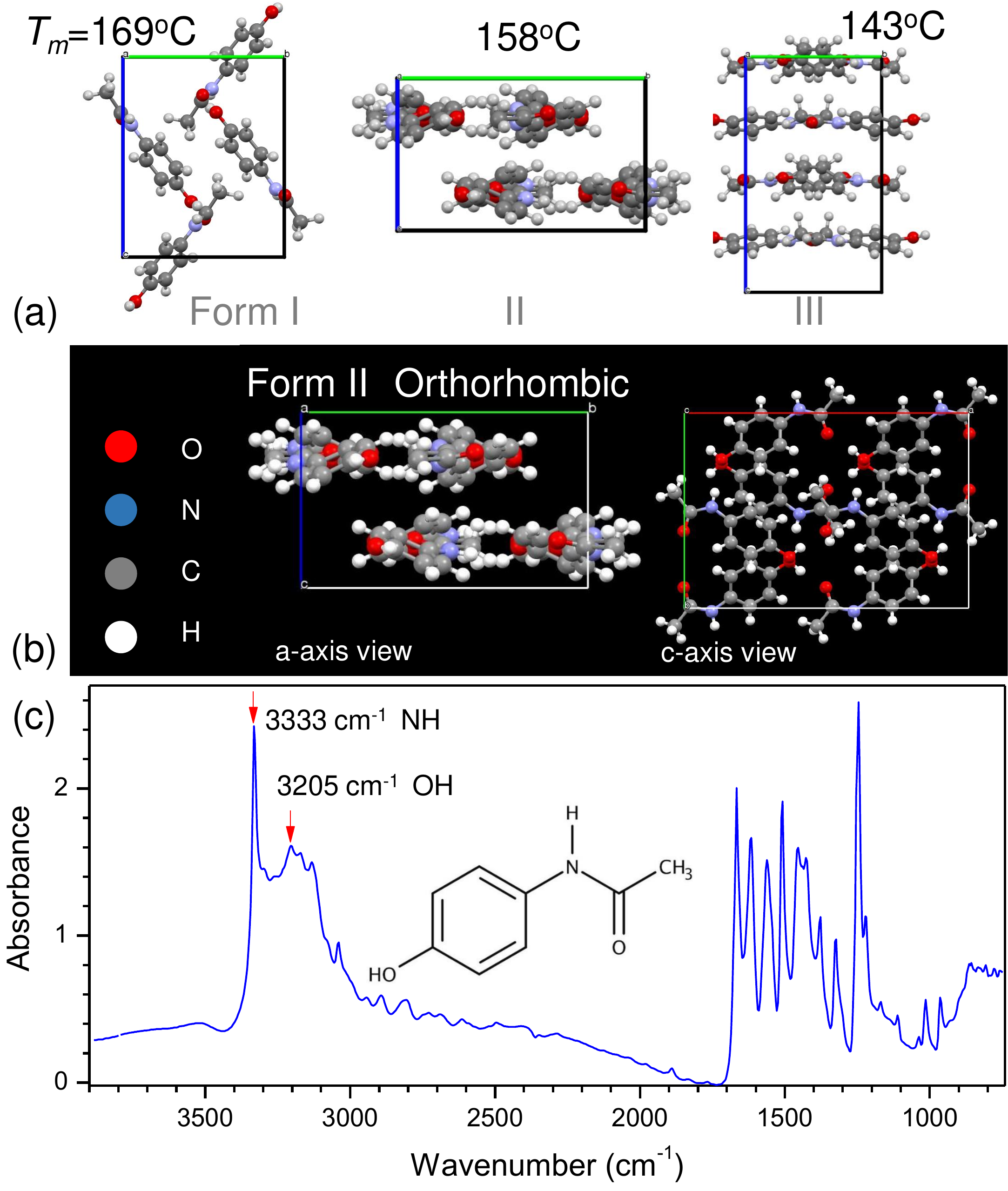}
\caption{(a) Molecular organisation \blue{of paracetamol} in \blue{the} forms I, II, III and their melting temperatures~\cite{Martino,Gaisford,Qi}. (b) Paracetamol form II crystal structure and (c)
the absorbance spectrum \blue{of the} form II~\cite{parac} averaged over the measured area at $\theta =
0^\circ$ (horizontal x-axis in transmission, $T$, measurements). The hydrogen bonding at 3205~cm$^{-1}$ was used for mapping and for orientation of the optical slow-axis at 3600~cm$^{-1}$, at which absorbance is close to zero and \blue{transmittance} is changing due to the retardance (see, Eqn.~\ref{e1}). The OH \blue{stretching} band is aligned with the molecular chain (inset in (c)), which is the \blue{optical} slow-axis.}
\label{f-parac}
\end{center}
\end{figure}
\begin{figure}[t!]
\begin{center}
\includegraphics[width=9.5cm]{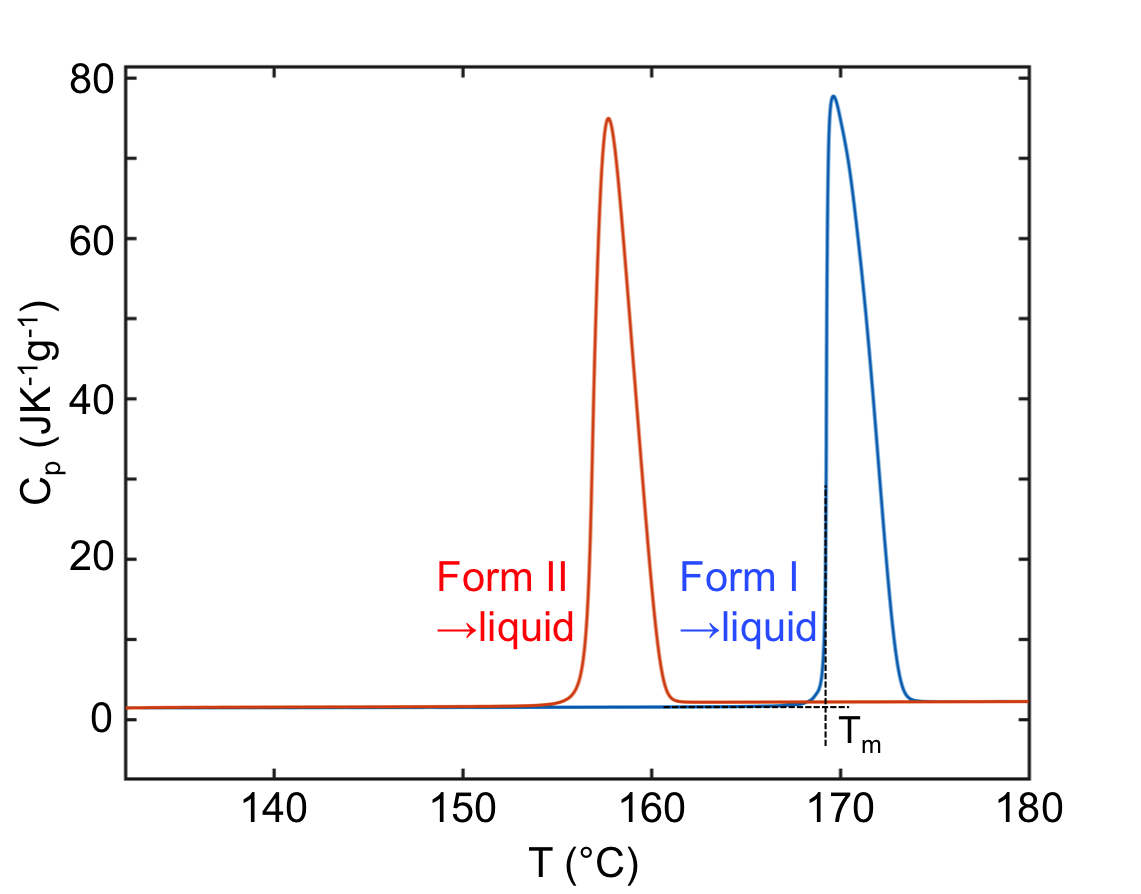}
\caption{\blue{Specific heat capacity $c_{\rm{p}}$  measured on heating at the rate of 10 K/min for the specimen as purchased (in the first heating, blue line) and after cooling at 10 K/min (in the second heating, red line). The highest temperature reached at the end of the first cycle was 180$^\circ$C.} }
\label{f-dsc}
\end{center}
\end{figure}
\begin{figure}[t!]
\begin{center}
\includegraphics[width=10cm]{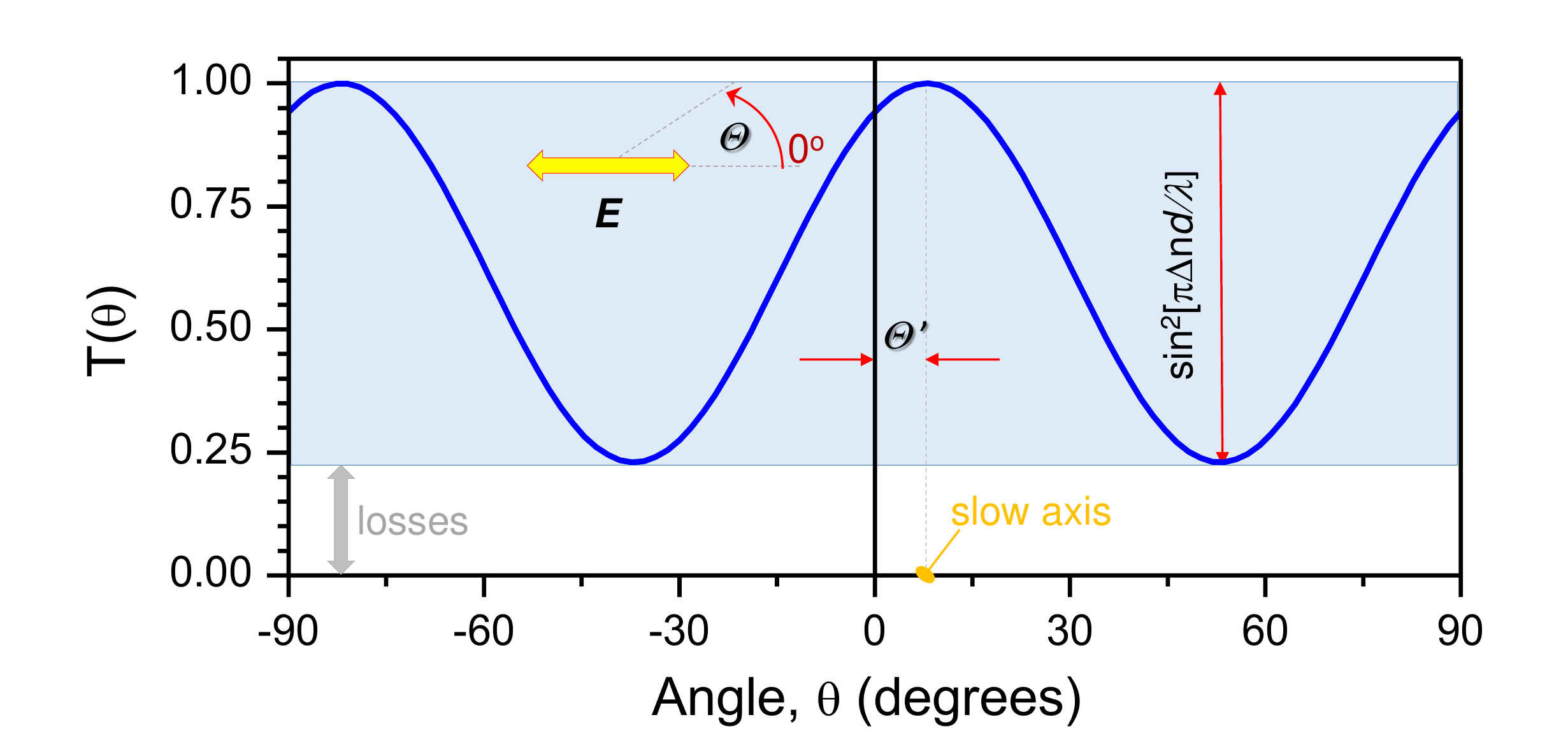}
\caption{Polarisation angle dependence of transmittance, $T(\theta)$ (Eqn.~\ref{e1}). Experimental transmittance is fitted by Eqn.~\ref{e1} at four polarisation angles~\cite{Hikima} $\theta$ separated by $\pi/4$; more angles increase \blue{a} fidelity of the fit. The marked losses represent collection and absorption losses measured through the parallel polariser - analyser setup. The slow\blue{-}axis orientation $\theta^{'}\equiv\theta^{'}_n$ (for pure retardance effect) and the retardance $\Delta nd/\lambda$ is also found from the fit. This measurement is made at selected wavelength.}
\label{f-T}
\end{center}
\end{figure}
\begin{figure}[ht!]
\begin{center}
\includegraphics[width=9cm]{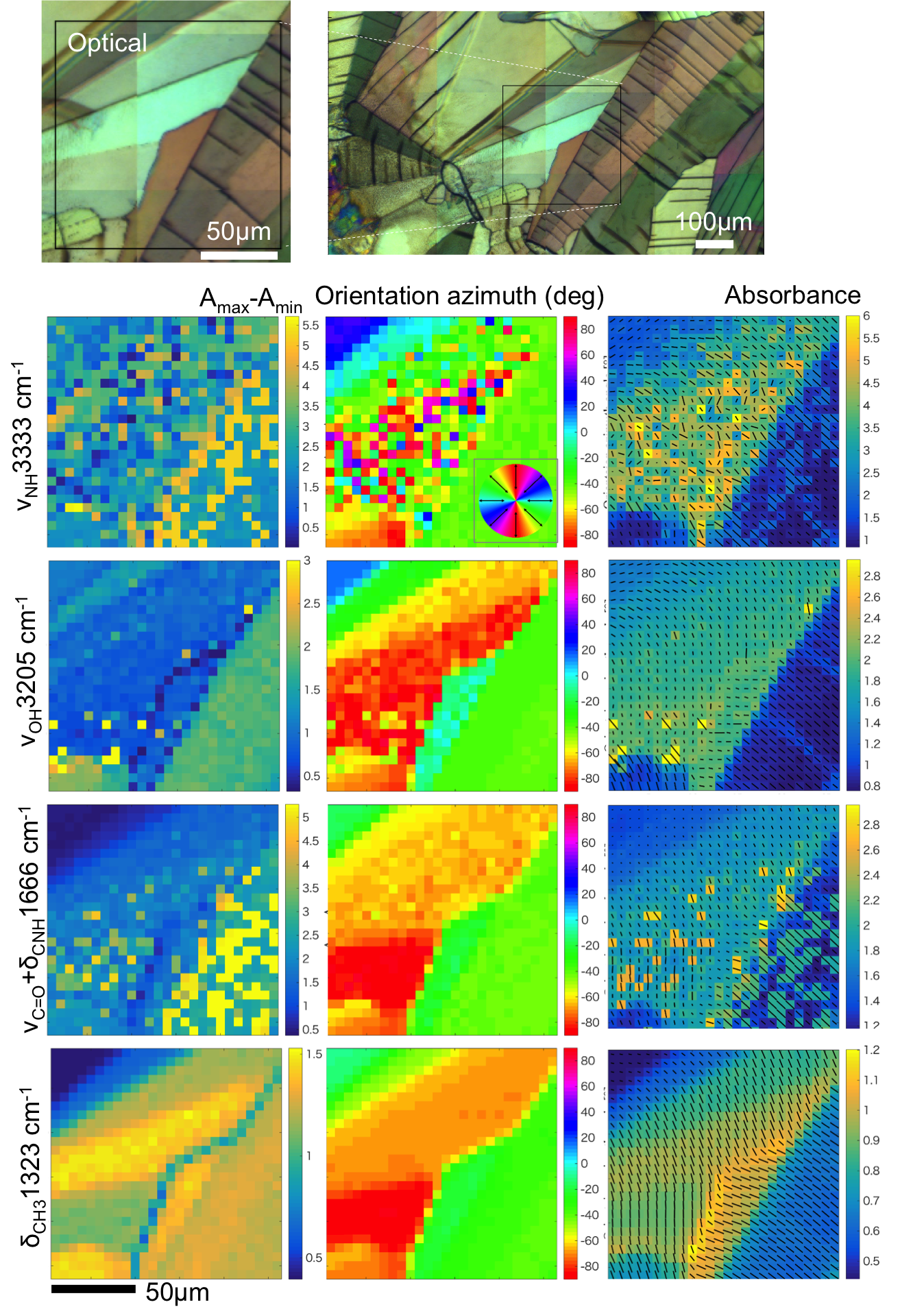}
\caption{Absorbance spectral mapping for amplitude and orientation for several specific molecular vibration modes (Fig.~\ref{f-parac})~\cite{Burgina}, where $\nu$ and $\delta$ define stretching and deformation, respectively. Top row shows an optical cross Nicol polariscopy images in a white light illumination. (left) The absorbance amplitude $A_{max}-A_{min}$ color map. (center) The orientation azimuth angle $\theta^{'}$ (Fig.~\ref{f-T}) color map. (right) The orientation vector map (black line) overlayed with the average absorbance color map, which was measured at $\theta = 0, 45, 90, 135^\circ$. Images for the orientation azimuth were taken at $\theta = 0, 15, 30, 45, 60, 75, 90, 135^\circ$  at 3333~cm$^{-1}$ and the length of the black line represents the orientation strength defined by amplitude $A_{max}-A_{min}$. The pixel pitch was 5~$\mu$m, optical resolution (projected aperture) 6.94~$\mu$m, spectral resolution $8$~cm$^{-1}$; synchrotron IR light source at the IR micro-spectroscopy beamline of the Australian synchrotron.}
\label{f-azim}
\end{center} 
\end{figure}

Here, optical spectroscopy \blue{based on} infrared (IR) molecular absorbance window was used to reveal the orientation of molecules in paracetamol. Absorbance and optical retardance (\blue{a change of} phase) related to \blue{the} real and imaginary parts of the refractive index $(n+i\kappa)$, respectively, were measured at different spectral windows and used to map micro-domains in the paracetamol form II. This was possible due to the parallel analyser-polariser setup used in this study since the ordinary IR imaging cannot measure the molecular orientation and the retardance simultaneously.  



\section{Samples and Procedures}\label{samp}
\subsection{Paracetamol}

Polymorphs of paracetamol~\cite{Haisa,parac,Perrin} and their thermal characteristics: the glass transition, $T_g$, and melting, $T_m$, temperatures and possible phase transitions at atmospheric pressure are shown in Fig.~\ref{f-forms}. Polariscopy images with a waveplate $\lambda = 530$~nm color shifter illustrates the structural morphology of the forms I, II, and III. \blue{The crystal structures of polymorphs of paracetamol are shown in Fig.~\ref{f-parac}.} Paracetamol \blue{(Sigma-Aldrich)} was melted between two 300-$\mu$m-thick CaF$_2$ plates and then quenched into amorphous phase. Then, the amorphous paracetamol was heated to 60~$^\circ$C and crystallized into the form III. The form III paracetamol was then heated to 110~$^\circ$C and transferred into the form II crystal which was used for \blue{the optical} characterisation in this study. \blue{For the differential scanning calorimetry (DSC) measurement, paracetamol in powder form (Sigma-Aldrich) was used without further purification and thermal treatment.} 　  

\begin{figure}[t]
\begin{center}
\includegraphics[width=6cm]{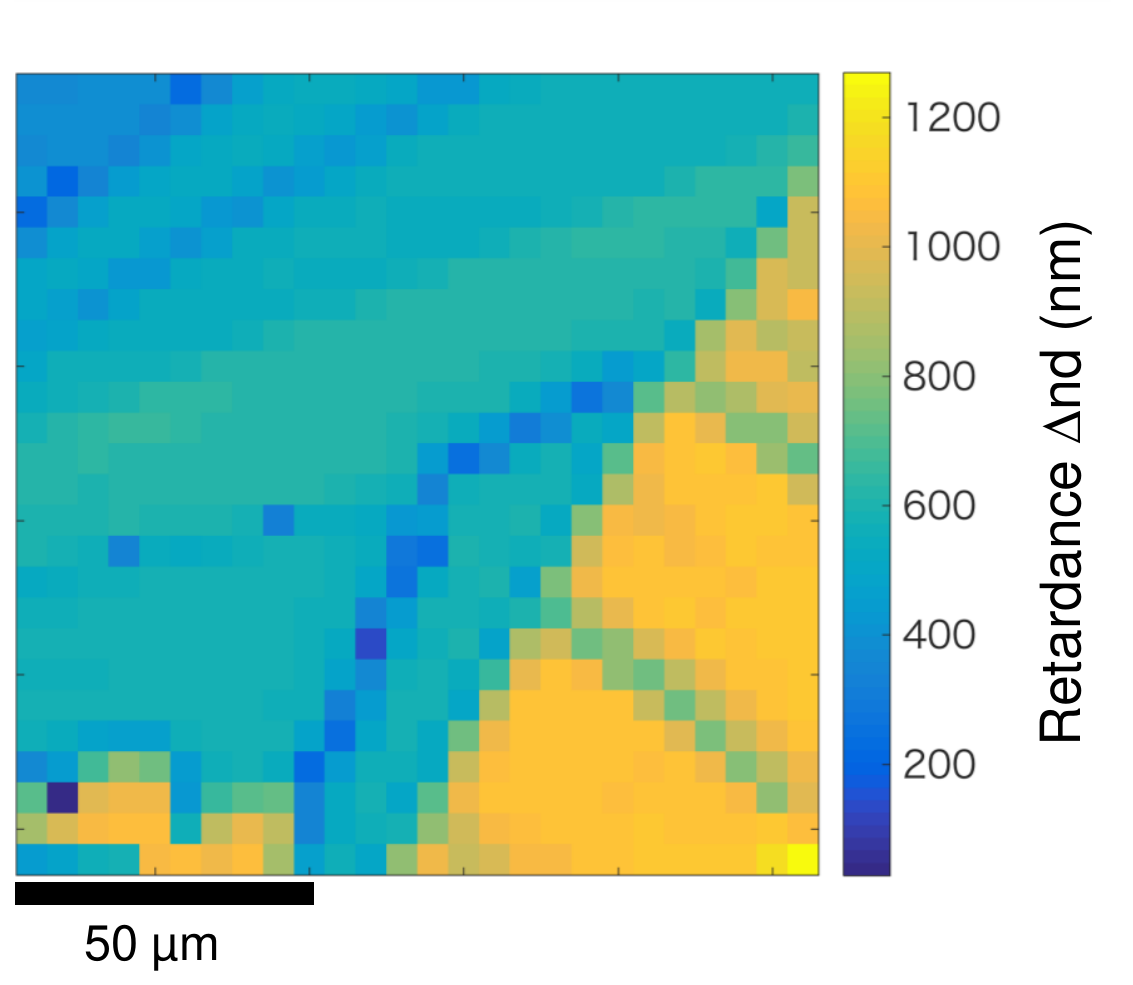}
\caption{Retardance $\Delta nd$ map of the same area shown in IR absorbance analysis in Fig.~\ref{f-azim} at 3600~cm$^{-1}$. The pixel pitch was 5~$\mu$m, optical resolution 6.94~$\mu$m; synchrotron IR light source at the IR micro-spectroscopy beamline of the Australian synchrotron. Retardance map was calculated as an average from measurements at $\theta = 0, 15, 30, 45, 60, 75, 90, 135^\circ$.}
\label{f-reta}
\end{center}
\end{figure}
\begin{figure}[t]
\begin{center}
\includegraphics[width=11.5cm]{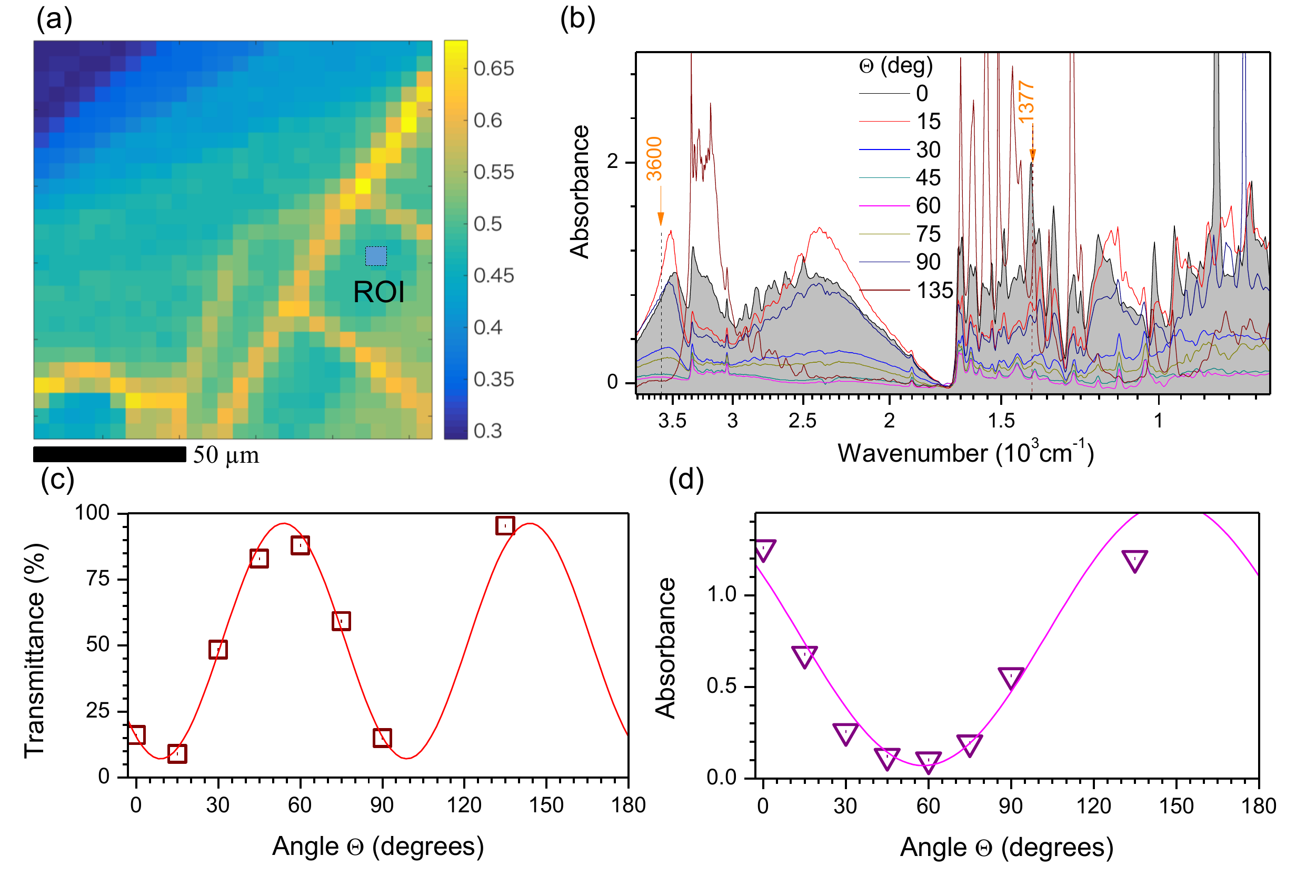}
\caption{(a) Color map of the spectral average of absorbance. ROI indicates the region-of-interest for single point spectral measurements;  ROI at $x=23$, $y=14$ when the upper left corner is (0,0).  (b) Single point spectra at different $\theta$ angles; note logarithmic abscise scale used to better separate \blue{absorption} bands. (c) Transmittance (Eqn.~\ref{e1}) change at non-absorption band of 3600 cm$^{-1}$. The best fit is plotted by $y=1-a\sin^2(2x-2b)+c$ which achieves regression coefficient $R^2 = 0.986$ with $a=0.893$, $b=-36.11$, $c=-0.036$. (d) Absorbance change at $\delta$(CH$_3$) band 1377~cm$^{-1}$ were also measured by transmission, however, at the \blue{absorption} band $A = -\lg(T)$ and Malus law applies due to the mutual orientation between the linear polarisation and orientation of the absorbing dipoles. The best fit was made by $y=a\cos(2x-2b)+c$ with $R^2 = 0.942$,  $a=0.72$, $b=-31.97$, $c=0.79$.
}
\label{f-all}
\end{center}
\end{figure}
\begin{figure}[t]
\begin{center}
\includegraphics[width=14cm]{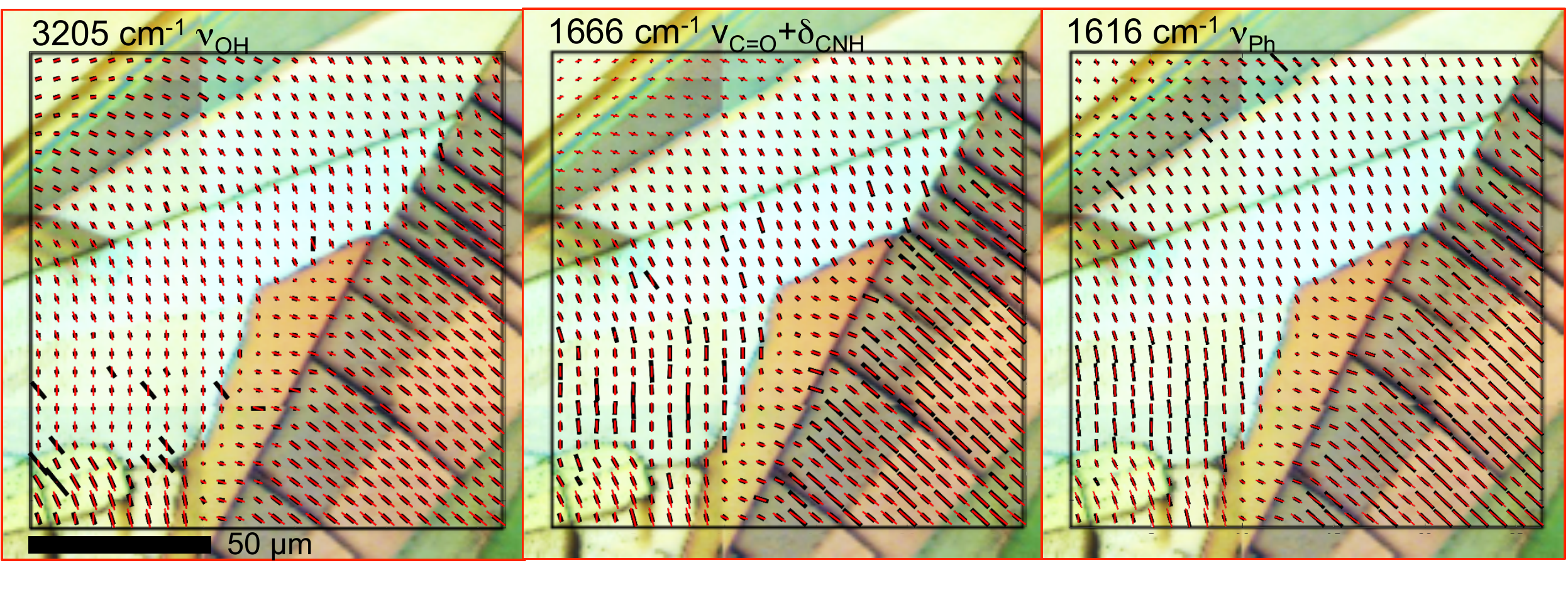}
\caption{Overlayed optical cross-polarised images of paracetamol form II and vector maps of orientation of the absorption $\theta^{'}$ (black-bars) at several absorbance bands with the optical slow-axis (retardance azimuth) $\theta^{'}_n$ (red-bars) measured at 3600~cm$^{-1}$ where is no absorption. The band $\nu_{Ph}$ is C=O, C=C stretching of the aromatic ring~\cite{Burgina}. See details of statistical analysis for the 3205~cm$^{-1}$ band in Sec.~\ref{suppl} Fig.~\ref{f-comp}.}
\label{f-both}
\end{center}
\end{figure}
 
\blue{\subsection{Differential scanning calorimetry}}

\blue{The thermal characterization was carried out by differential scanning calorimetry (Perkin Elmer DSC 7). The sample of a few milligrams of paracetamol was placed in an aluminum pan for measurements at temperature scan speed of 10~K/min. The calibration of temperature was made based on the melting point of indium while for the specific heat capacity by using sapphire. First heating scan was performed on the purchased paracetamol to 180$^\circ$C, held at that temperature for 3~min, and cooled at 10~K/min to 20~$^\circ$C. After a 3~min pause, a second heating scan was performed. The starting point of the endothermic peak was taken as the melting point $T_m$ (black dotted line in Fig.~\ref{f-dsc} ), and the peak area was taken as the melting enthalpy $\Delta H_m$ of each polymorph phase.}

\subsection{Spectral mapping of transmittance}

Optical characterisation was carried out by measuring transmittance of the \blue{sample} with parallel polariser-analyser. For transparent sample when reflectance, $R$, and absorbance, $A$,  are negligible, the polarisation dependence of transmittance through the parallel  analyser-polariser is given by:
\begin{equation}\label{e1}
T(\theta) = 1 - \sin^22(\theta-\theta^{'}_n)\sin^2(\pi\Delta n
d/\lambda),
\end{equation}
\noindent where $d$ is the thickness of the sample, $\Delta n$ is the birefringence, $\lambda$ is selected wavelength at which $T$ is measured, and $\theta$ is the polarisation angle ($0^\circ$ corresponds to horizontal x-axis and it is positive for the anti-clockwise rotation). Note, the second term in Eqn.~\ref{e1} represents the transmittance through the crossed polariser-analyser. Figure~\ref{f-T} shows the polarisation dependence of transmittance (Eqn.~\ref{e1}) and determination of the angle of slow (or fast) optical axis, $\theta^{'}_n$; there is an uncertainty between the slow- and fast-axis determination. The usual convention that the strongest absorption orientation is set as the slow-axis \blue{was used}. The minimum number of points for the fit (Fig.~\ref{f-T}) is three, usually, separated by $\theta = \pi/4$ due to the three unknown fitting parameters of $\sin$-wave: amplitude, phase, and initial phase angle $\theta^{'}_n$; a larger number of points increase fidelity of the fit and was implemented in this study. In addition to the determination of $\theta^{'}_n$ from the fit, the retardance $\Delta n d/\lambda$ is also obtained from the amplitude of the $\sin$-fit (see Fig.~\ref{f-T}).

In the case of absorbance $A = -\lg T$ the fitting function was~\cite{Hikima}: 
\begin{equation}\label{e2}
A(\theta) = \frac{A_{max} - A_{min}}{2}\cos2(\theta - \theta^{'})+\frac{A_{max} + A_{min}}{2},
\end{equation}
where $A_{max,min}$ are the maximum and minimum absorbance and $\theta^{'}$ is the orientation of the maximum absorbance (dipoles aligned with linear polarisation of the incident light).
   
IR spectral mapping \blue{measurement} of absorbance and retardance was carried out by measuring transmittance using \blue{a Bruker Vertex 80v spectrometer coupled with a Hyperion 2000 FTIR microscope and a liquid nitrogen-cooled narrow-band mercury cadmium telluride (MCT) detector (Bruker Optik GmbH, Ettlingen, Germany). } The polarizer was ZnSe wire-grid (Edmund). Light source was \blue{synchrotron IR beam at} the IR micro-spectroscopy  beamline of the Australian synchrotron.

Multi-parameter maps of selected areas were measured close to the wavelength resolution at several orientation angles. The orientation of absorbing species as well as optical slow/fast-axis were calculated. Overlayed multi-parameter maps were made for specific wavelengths from the hyper-spectral FTIR data.  

\section{Results and Discussion}

\blue{Thermal analysis was carried out to characterize the thermodynamic stability of each polymorph, as shown in Fig.~\ref{f-dsc}. The endothermic peak due to  the melting of the form I crystal at $T_m$ = 169$^\circ$C and $\Delta H_m$ = 185~J/g were observed during the first heating. In the second heating, the endothermic peak was observed, which was due to the melting from the form II to liquid at $T_m$ = 157$^\circ$C and the $\Delta H_m$ =  173~J/g. These results are equivalent to literature values~\cite{Boldy,Klim}.  The thermodynamic unstability of form II crystal is explained by the lower $T_m$ and $\Delta H_m$ than form I. It also suggests a higher water solubility~\cite{Bhattachar}.}

\blue{In the IR spectral mapping,} we focus on the form II phase of paracetamol since it \blue{has the higher water solubility than} form I. Films of $\sim 10~\mu$m thickness were prepared for optical IR spectroscopic analysis (see top row of Fig.~\ref{f-azim}) as described in Sec.~\ref{samp}. By using the described method of FTIR mapping of spectral transmittance, it is possible to determine the absorbance at the specific wavenumber (Fig.~\ref{f-parac})~\cite{Burgina} and calculate  the orientation of the absorbing species by the model used in this study (Fig.~\ref{f-T}). It is based on the four-angles method~\cite{Hikima}. By measuring absorbance $A=-\lg T$ at several polarisation angles, the molecular dipoles contributing to the absorbance by the Malus law $\propto\cos^2\theta$ depending on the absorbing dipole orientation can be calculated from the fit (Fig.~\ref{f-T}). 

The orientation of the absorbing species can be obtained as summarised in Fig.~\ref{f-azim} for four specific lines of NH stretching ($\nu$) vibration at 3333~cm$^{-1}$, OH at 3205~cm$^{-1}$, C=O stretching coupled with CNH deformation $\delta$ at 1666~cm$^{-1}$ and CH$_3$ deformation at 1323~cm$^{-1}$. Wealth of information can be retrieved from the presented maps revealing ordered and disordered regions. Optical polariscopy image (top of Fig.~\ref{f-azim}) reveals birefringence by color, i.e., the retardance $\Delta nd/\lambda$ at the visible wavelengths with clearly defined domain boundaries (dark lines) between the differently oriented regions in the paracetamol form II. The IR absorbance and azimuthal orientation of the particular vibrational bands also reveals the orientational dependencies, presence of disorder and grain boundaries (see the CH$_3$ deformation at 1323~cm$^{-1}$ in Fig.~\ref{f-azim}). Since an oversampling was used in image acquisition, i.e. the optical resolution with the chosen pinhole was $\sim 7~\mu$m, while the step size was only $5~\mu$m, the domain boundaries are blurred. 

For the retardance at the IR wavelengths we have chosen 3600~cm$^{-1}$ wavenumber (or 2.78~$\mu$m wavelength) where there is no absorbance. \blue{The retardance image is shown in Fig.~\ref{f-reta}.} The orientation azimuth $\theta^{'}_n$  at this wavelength was selected as the optical slow-axis (note, it is determined for the real part of the refractive index (Eqn.~\ref{e1}) to make a distinction from the $\theta^{'}$ reserved for the anisotropic absorbance bands). Next, the orientation of the slow\blue{-}axis at 3600~cm$^{-1}$ wavenumber is compared with orientation of absorbing dipoles at several specific bands with details shown in Sec.~\ref{suppl}.  
Figure~\ref{f-all} shows area mapping and orientation determination of the optical slow(fast)-axis in the case of non-absorbing spectral position at chosen 3600~cm$^{-1}$ and at the absorption band at 1377~cm$^{-1}$. The measurements were carried out at a single pixel level (ROI is selected on the spectral map of absorbance (a)). A strong angular anisotropy of absorbance (transmittance) are shown in Fig.~\ref{f-all}(b) for the eight orientation angles $\theta$. The transmittance was directly fitted with Eqn.~\ref{e1} $y=1-a\sin^2(2x-2b)+c$ and achieves a good fit with a regression coefficient $R^2 = 0.986$ (Fig.~\ref{f-all}(c)). The parameter $b$ of the fit directly defines the orientation angle $\theta^{'}_n$, which is denoted with $n$ to distinguish the case of a pure retardance effect caused by the real part of the refractive index. For the absorption band 1377~cm$^{-1}$, the fit to the Malus law $\propto\cos^2\theta$ was chosen in an equivalent form $y=a\cos(2x-2b)+c$ (Fig.~\ref{f-all}(d)). In this form  $a =(A_{max}-A_{min})/2$, $c=(A_{max}+A_{min})/2$, and the strongest absorption is defined by the orientation angle $b=\theta^{'}$ (Eqn.~\ref{e2}).

From the fits presented in Fig.~\ref{f-all}, it is possible to establish the orientation angles $\theta^{'}$ and $\theta^{'}_n$ for the absorbing species and optically retarding (non-absorbing) spectral regions, respectively. Also, the amplitudes as $A_{max}-A_{min}$ for absorbance or transmittance can be calculated for each pixel (see details in Sec.~\ref{suppl} for presentation in the format of vector plots). 

Finally, the vector plots of orientation and amplitude of the absorbance and transmittance (in non-absorbing regions of spectrum)  at IR wavelengths can be compared and referenced to the optical images which have better resolution. The data are summarised in Fig.~\ref{f-both}. Orientation of the maximum absorption is correlated with that for retardance with \blue{the} largest uncertainty at the domain boundaries. Here, we introduce the method of analysis, which has to be explored deeper with a more dedicated study to reveal its potential. \blue{An overestimation of birefringence in thinner regions of the structure are expected as we demonstrated in the silk fiber~\cite{18sr}.} 

It should be possible to use the presented analysis for the anomalous and normal dispersion regions near absorption bands to separate absorption and retardance contributions to transmittance. Such information is currently not experimentally accessible and could help a limited understanding of phase transitions. This method can be used for analysis of bio-materials, e.g., laser structured dominantly crystalline silk fibers and its amorphisation into water soluble form~\cite{17mre115028}, molecular orientation in micro-thin silk fibers~\cite{17sr7419} at different spectral ranges~\cite{17m356}, and bactericidal bio-materials~\cite{16l4698} where structure influence to the biocidal function has to be better understood. Internal laser structuring of transparent glasses and dielectrics~\cite{02ass705p} can also benefit from the presented analysis of optical anisotropy in real and imaginary parts of the refractive index.

The uncertainty in determination of the slow\blue{-} or fast\blue{-}axis (at visible wavelengths) can be removed by the larger number of measurement points for the fitting using an additional electrically tunable liquid crystal (LC) retarder which enables to scan reference retardation $\theta^{LC}_n$ continuously. 
From observation of transmittance, one can  determine which axis is slow $n_e$ and fast $n_o$ in the measured sample~\cite{18sr}.

\section{Conclusions and outlook}

Optical mapping of paracetamol form II was carried out at the wide range of IR spectrum at the absorbing and non-absorbing (where light  experienced a pure optical retardance) spectral domains. Using angular dependence of transmittance $T$ it is possible to determine angular orientation of optically absorbing and retarding moieties. The presented vector and color plot analysis is a useful tool to combine and compare data at different spectral regions.   

With presented vector plot analysis it should be possible to monitor orientation changes during phase transitions in real time using array mapping with bright synchrotron radiation.

\section*{Acknowledgements}
JM acknowledges a partial support by a JSPS KAKENHI Grant No.16K06768 and No. 18H04506, SJ was supported via ARC Discovery DP170100131 and PLASENS Consortium FNR C15/MS/10459961 projects. Experiments were carried out through the beamtime proposals ID. 12107 and 13416 at the Australian Synchrotron IRM Beamline at the Australian synchrotron, part of ANSTO. 

\small{\section*{References}

\appendix 
\section{Supplement}\label{suppl}
\setcounter{figure}{0}    

Figure~\ref{f-comp} shows presentation of data which was used for vector plots. Using Eqn.~\ref{e1} the best fit of transmittance was achieved for the eight sample orientation angles $\theta$ and the orientation of optical slow-axis $\theta^{'}_n$ was determined (Fig.~\ref{f-T}). At the absorption band 3205~cm$^{-1}$ the transmission was used to calculate absorbance $A=-\lg T$ and was fitted with Malus law function. The orientation $\theta^{'}$ was obtained together with the amplitude of the absorbance. Panel (b) shows vector plot of the orientation and amplitude shown for each pixel. Correlation between the orientations of $\theta^{'}_n$ and $\theta^{'}$ for all pixels is shown in panel (c).    

\begin{figure}[tbh]
\begin{center}
\includegraphics[width=14cm]{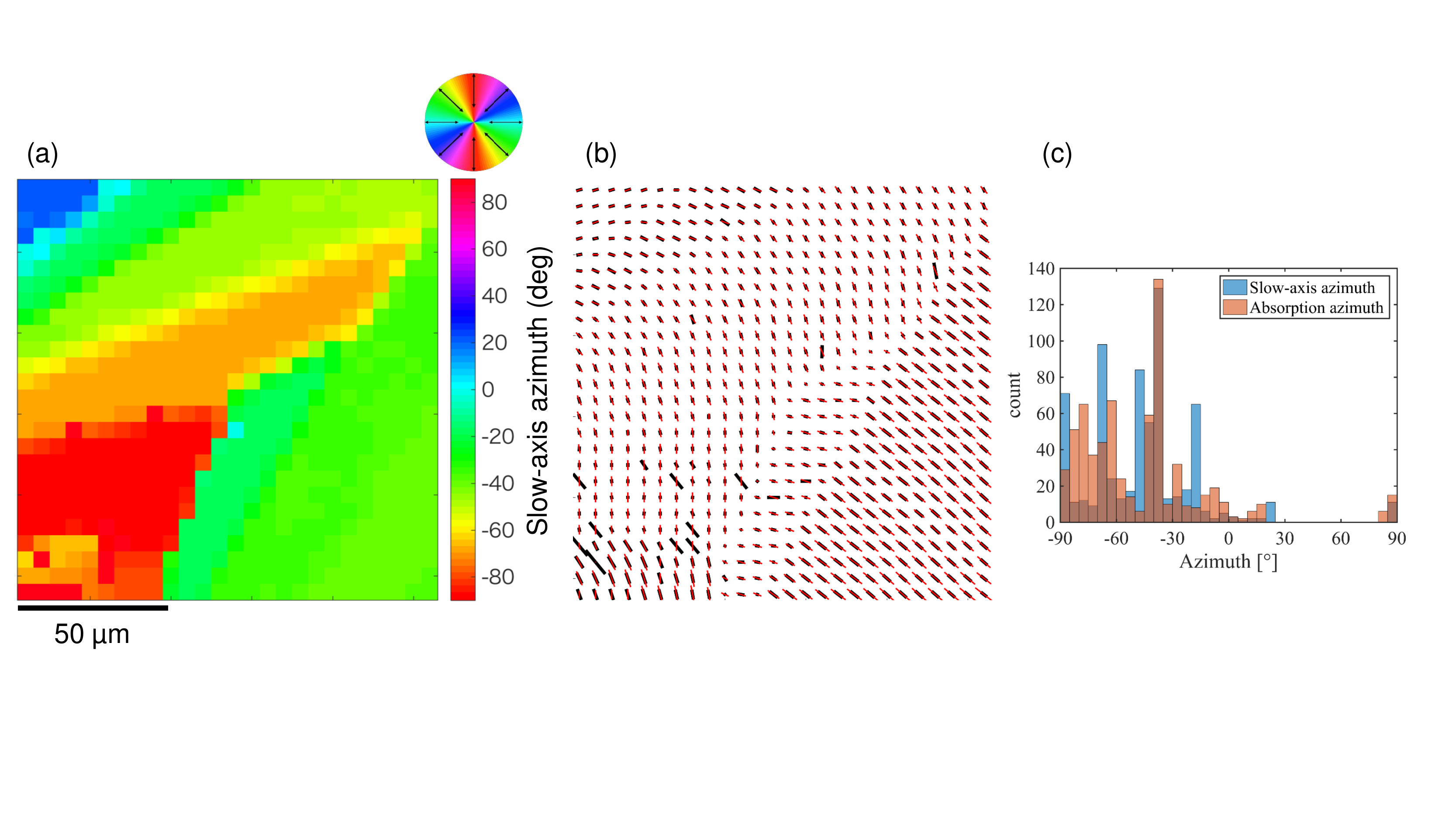}
\caption{(a) Slow-axis azimuth angle $\theta^{'}_n$ color map at 3600~cm$^{-1}$. Since there is ambiguity in determination between slow- and fast-axis ($90^\circ$ degrees difference) we used orientation of the OH band at 3205~cm$^{-1}$ as slow-axis (see the molecular structure in the inset of  Fig.~\ref{f-parac}(c)). (b) Black-bar represents the  $\nu_{OH}$ at 3205~cm$^{-1}$ absorption orientation $\theta^{'}$ vector map, the red-bar at 3600~cm$^{-1}$ slow-axis vector map (the amount of retardance is plotted in Fig.~\ref{f-reta}). (c) The orientation azimuths  $\theta^{'}_n$  and  $\theta^{'}$. The darkest sections of the bars are representing a mutual overlap between the orientation of the absorbance and retardance.}
\label{f-comp}
\end{center}
\end{figure}

\end{document}